\documentclass[prl,twocolumn,superscriptaddress,altaffillsymbol]{revtex4-1}
\usepackage{amssymb}
\usepackage{amsmath}
\usepackage{commath}
\usepackage{graphicx}
\usepackage{verbatim}
\usepackage{epsfig}
\usepackage{epstopdf}
\usepackage{color}
\usepackage{siunitx}
\usepackage{array}
\usepackage{upgreek}
\newcolumntype{P}[1]{>{\centering\arraybackslash}p{#1}}
\newcolumntype{M}[1]{>{\centering\arraybackslash}m{#1}}

\DeclareSIUnit\gauss{G}

\begin{document}
	
\title{Nodeless bulk superconductivity in the time-reversal symmetry breaking Bi/Ni bilayer system}

\author{Prashant Chauhan}
\affiliation{The Institute for Quantum Matter, Department of Physics and Astronomy, Johns Hopkins University, Baltimore, Maryland 21218, USA}

\author{Fahad Mahmood}
\email{fahad@jhu.edu}
\affiliation{The Institute for Quantum Matter, Department of Physics and Astronomy, Johns Hopkins University, Baltimore, Maryland 21218, USA}

\author{Di Yue}
\affiliation{Department of Physics, Fudan University, Shanghai, China}

\author{Peng-Chao Xu}
\affiliation{Department of Physics, Fudan University, Shanghai, China}

\author{Xiaofeng Jin}
\affiliation{Department of Physics, Fudan University, Shanghai, China}

\author{N.~P.~Armitage}
\email{npa@jhu.edu}
\affiliation{The Institute for Quantum Matter, Department of Physics and Astronomy, Johns Hopkins University, Baltimore, Maryland 21218, USA}

%\date{\today}

\begin{abstract}
	Epitaxial bilayer films of Bi(110) and Ni host a time-reversal symmetry (TRS) breaking superconducting order with an unexpectedly high transition temperature $T_c = \SI{4.1}{\kelvin}$. Using time-domain THz spectroscopy, we measure the low energy electrodynamic response of a Bi/Ni bilayer thin film from $\SI{0.2}{\tera\hertz}$ to $\SI{2}{\tera\hertz}$ as a function of temperature and magnetic field. We analyze the data in the context of a BCS-like superconductor with a finite normal-state scattering rate. In zero magnetic field, all states in the film become fully gapped, providing important constraints into possible pairing symmetries. Our data appears to rule out the odd-frequency pairing that is natural for many ferromagnetic-superconductor interfaces. By analyzing the magnetic field-dependent response in terms of a pair-breaking parameter, we determine that superconductivity develops over the entire bilayer sample which may point to the $ p $-wave like nature of unconventional superconductivity.
\end{abstract}

\maketitle
\setlength\belowcaptionskip{-3ex}

Unconventional superconductors that break time reversal symmetry (TRS) are promising platforms to realize Majorana edge modes. A remarkable candidate is a Bi(110) thin film deposited on a ferromagnetic Ni layer. This Bi/Ni bilayer system can have a  $T_c $ as high as $ \SI{4.1}{\kelvin}$  \cite{Moodera_PRB_1st_1990,LeClair_PRL_2005}, which is quite unexpected for a number of reasons. Elemental bismuth (Bi) has a high atomic mass and low Fermi energy/velocity; factors which generally preclude superconductivity according to standard BCS theory. Similarly, Ni is not superconducting at any temperature and, within conventional models of superconductivity, its ferromagnetism should inhibit rather than enhance superconductivity in the adjoining Bi layer \cite{Ginzburg_JETP_1956,Thomas_1969}.
\par With advances in epitaxial film growth and developments in topological and TRS breaking superconductivity, there has been renewed interest in this Bi/Ni bilayer system \cite{Gong_Sci_ad_2017, Gong_chin_phylett_2015, Gong_arxiv_2015, ZHOU_2017, WANG_PRB_2017, KESKIN_2017}.  Two key results include the observation of a zero-bias anomaly in point-contact Andreev reflection \cite{Gong_chin_phylett_2015}, a possible indicator of Majorana modes, and broken TRS as determined by polar Kerr effect measurements \cite{Gong_Sci_ad_2017}. TRS breaking suggests a complex pairing symmetry such that the phase of the superconducting order parameter winds around the Fermi surface. Examples of complex pairing include $d_{xy}\pm id_{x^2-y^2}$ which corresponds to even parity pairing and $p_{x}\pm ip_y$ which consists of odd parity pairing. Since this system is non-centrosymmetric and has large spin-orbit coupling, the superconducting order may be a novel pairing state with a mixture of even and odd parity components \cite{Bauer_2012,Schenck_1985}.

\begin{figure*}
	\includegraphics[width=\textwidth,height=4.2cm]{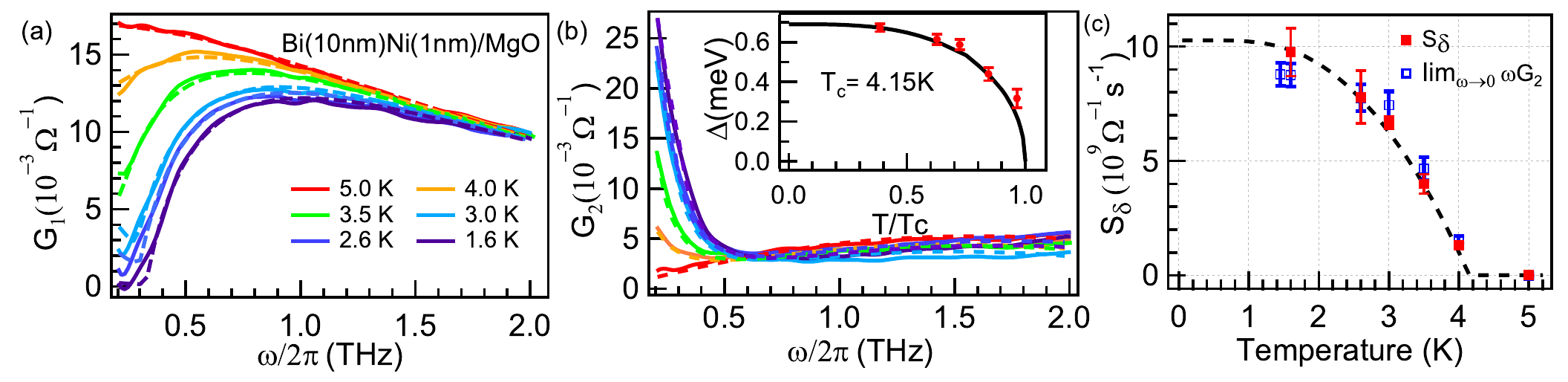}
	\caption{\label{fig:zerofield} (a) Real and (b) imaginary parts of the zero-field complex conductance of sample-A as a function of frequency from ($\SI{5}{\kelvin}>T_c$) to ($\SI{1.6}{\kelvin}<<T_c$) with fits (dashed lines) to the data using Mattis-Bardeen theory for a BCS superconductor with a finite normal state scattering rate. Inset: Extracted temperature dependent energy gap with fit (Solid line) to a BCS superconductor in the weak coupling limit. (c) Temperature dependent superfluid spectral weight, $ S_\delta $. Red squares show the difference between the spectral weights of $G_1(\omega)$ at $ \SI{5}{K} $ and various temperatures below $T_c$. Blue squares show $\lim_{\omega \to 0} \omega G_2$. The dashed line is the predicted superfluid spectral weight for a weakly coupled BCS superconductor. The error bars represent 2 s.d.}
\end{figure*}

\par There are two natural questions associated with the unconventional superconductivity in this system: (1) what is the gap structure of the superconducting order and does it have nodes or not? And (2) what is the mechanism for the superconductivity and where does it develop? Addressing these questions can have profound implications for the pairing symmetry in this system. For instance, it was proposed in Ref.~\onlinecite{Gong_Sci_ad_2017} that this system exhibits $d_{xy}\pm id_{x^2-y^2}$ superconductivity as it is the lowest angular momentum state which is TRS violating, consistent with strong spin-orbit coupling and the approximate surface symmetries of this system. This proposal is based on superconductivity occurring on the Bi surface opposite to the Bi/Ni interface, as suggested by a systematic study of the thickness dependence of each of the Bi and Ni layers \cite{Gong_chin_phylett_2015}. On the other hand, a few studies \cite{Liu_PRM_2018, Siva_JAP_2015,SIVA_JAP_2016} suggest that superconductivity occurs in the bulk of the system (perhaps due to the presence of s-wave superconducting alloys such as NiBi$_3$ which may occur due to diffusion in the Bi/Ni interface). It was proposed recently \cite{Chao_arxiv_2018} that this form of superconductivity combined with strong spin-orbit coupling of the Bi layer and the in-plane magnetic field of the Ni layer can lead to an effective $p_x \pm ip_y$ superconductivity instead of $d_{xy}\pm id_{x^2-y^2}$.

\par Here we use time-domain THz spectroscopy (TDTS) to systematically study and track the superconducting gap as a function of both temperature and magnetic field. We find the gap is nodeless and can be described phenomenologically in terms of a weakly coupled BCS theory. Analysis of the field-dependent optical conductance, points to superconductivity developing in the entire bilayer and not just the top surface. Moreover, from the calculation of the Fermi velocity of the superconducting charge carriers, it appears that superconductivity does not develop in either the Bi or Ni electronic states independently.

\par
A $\SI{10}{\nano\meter}$ thick rhombohedral Bi(110) layer was epitaxially grown on a $\SI{1}{\nano\meter}$ Ni(100) layer at $\SI{110}{\kelvin}$, which is seeded on a $\SI{0.5}{\milli\meter}$ thick MgO(100) substrate at $\SI{300}{K}$. TDTS measurements were performed on a total of three samples, each with the same T$_c$ of $ \SI{4.15}{K} $. They all gave similar results except for small differences at the lowest frequencies which may be due to differences in disorder levels. Both the real and imaginary parts of the complex conductance, $\tilde{G}(\omega)$, were obtained from the TDTS measurements, performed down to $\SI{1.6}{\kelvin}$ in both in-plane and out-of-plane magnetic fields.

\par
Fig.~\ref{fig:zerofield}(a-b) shows the temperature dependent $\tilde{G}(\omega)$ of the Bi/Ni bilayer between $0.2-\SI{2}{\tera\hertz}$ at zero magnetic field. In the normal state ($\SI{5}{K}$) the real part of $\tilde{G}$, $G_1(\omega)$, shows a Drude-like Lorentzian peak feature while the imaginary part, $\tilde{G}_2(\omega)$ shows a positive dispersion corresponding to a finite scattering rate. We model the normal state data using a Drude-Lorentz for $\tilde{G}(\omega)$ (SM). From the fit, $G_1(\omega)$ in the limit $\omega \to 0$ is found to be $ \SI{17.0}{\SIUnitSymbolOhm^{-1}} $, which matches quite well with the dc conductance measurement, $G_{dc} = \SI{17.4}{\SIUnitSymbolOhm^{-1}}$ (SM). It is important to point out that the normal state conductance of the bilayer is far larger than layers of just Bi(110) or Ni(001) individually (SM) showing that the electronic structure of the bilayer is different than either of these materials. Below $T_c$, both $G_1(\omega)$ and $G_2(\omega)$ show features indicative of a fully gapped superconductor. As the temperature falls below $T_c$, a strong depletion develops in $G_1(\omega)$ (solid lines Fig.~\ref{fig:zerofield}a) at low $\omega$, corresponding to the opening up of the superconducting gap. The small $G_1(\omega)$ at sub-gap frequencies is due to the contribution of thermally excited quasi-particles, which becomes exponentially small as the temperature is lowered. Quite interestingly \textit{all} metallic carriers appear to become gapped; to within our experimental sensitivity there is no remnant metallic layer that does not go superconducting. This is also clear from a comparison of this data with the measured $G_1(\omega)$ of just Bi(110) and just Ni(001) individually (SM). $G_2(\omega)$  (Fig.~\ref{fig:zerofield}b), increases as $\omega \to 0$ for T $<$ $ T_c $ and shows a 1/${\omega}$-like dependence at the lowest temperatures and frequencies, characteristic of the superconducting state. 
\par
To determine the superconducting gap $\Delta$, we simultaneously fit $G_1(\omega)$ and $G_2(\omega)$ using Mattis-Bardeen theory \cite{Mattisbardeen_PR_1958,Tinkham_PR_1956,Tinkham_PR_1968} for a uniformly gapped superconductor with a finite normal-state scattering rate (SM)\cite{ZIMMERMANN_PhyC_1991}. For the fitting procedure, the only free parameter is the superconducting gap, $\Delta$(T), while the scattering rate and the plasma frequency are kept fixed to the values determined from the normal-state $G_1(\omega)$, as discussed above. The results of the fits are shown as dashed line in Fig.~\ref{fig:zerofield}(a-b). The fit at the lowest temperature gives, $\Delta$($\SI{1.6}{\kelvin}$) = $\SI{0.67}{\milli\electronvolt}$ which is similar to the value obtained from tunneling spectroscopy ($\SI{0.64}{\milli\electronvolt}$) \cite{LeClair_PRL_2005}.
\begin{figure*}
	\includegraphics[width=17.80cm]{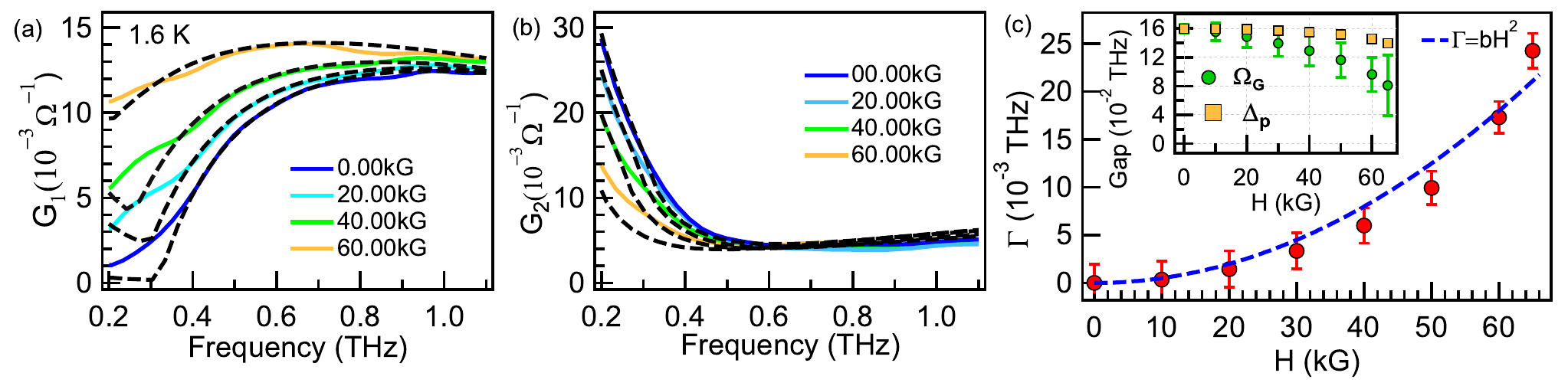}
	\caption{\label{fig:Parallel_MB}(a-b) In-plane field dependent real, $G_1(\omega)$, and imaginary part $G_2(\omega)$ (solid lines) of the complex conductance for Bi/Ni bilayer sample-B at $\SI{1.6}{\kelvin}$ with fits (dashed lines) modeled using Mattis-Bardeen theory for an effective spectroscopic gap, $\Omega_G$. (c) Field dependence of pair-breaking parameter $\Gamma$, determined from optical conductance along with fit $ \Gamma = bH^2 $ (dashed line). Inset: the field dependence of $\Omega_G$ (green dots) and pair-correlation gap $ \Delta_p $ (orange squares) for the Bi/Ni bilayer. The error bars represent the $\SI{95}{\percent}$ confidence interval.}
\end{figure*}
The close agreement between the experimental data and Mattis-Bardeen fits indicates that the electrodynamic response of the Bi/Ni bilayer system below $T_c$ corresponds to that of a fully gapped superconductor. From the fitting, we obtain the zero temperature gap as 2$\Delta$(0) = $\SI{0.334}{\tera\hertz}$ ($\SI{1.38}{meV}$) or 2$\Delta(0)/k_B T_c=3.85$, i.e., very close to the weak coupling limit of $3.53$ for a fully gapped BCS superconductor.  The temperature evolution of the superconducting gap, $\Delta$(T), (inset of Fig.~\ref{fig:zerofield}b) closely follows the expected form for a BCS superconductor in the weak-coupling limit, as given by the standard numerical approximation $\Delta(T) =\Delta(0)\tanh{[1.74\sqrt{\textstyle T_c/T-1}]}$ (black line). The observation of fully gapped superconductivity appears to rule out odd-frequency pairing that is natural for ferromagnetic-superconductor interfaces. Odd-frequency pairing is expected to have subgap spectral features \cite{linder2009pairing,tanaka2007theory,bergeret2005odd,pal2017spectroscopic}.
\par
To confirm the Mattis-Bardeen fits and get further insights into the superconducting gap structure, we study the temperature dependence of the superfluid spectral weight ($S_\delta$) as a direct measure of the superfluid density. Using the Ferrel-Glover-Tinkham (FGT) sum rule, $S_\delta$ can be extracted through $S_\delta = S_n - S_{qp}$, where $S_n$, the total spectral weight, is determined by the area under the $G_1(\omega)$ curve for the normal state Drude conductance at $\SI{5}{K}$ and $S_{qp}$, the quasi-particle spectral weight, is the area under the $G_1(\omega)$ curve for temperatures below $T_c$. It can be seen in Fig.\ref{fig:zerofield}c, the temperature evolution of $S_\delta(T)$ extracted using this method follows the predicted behavior of a fully gapped BCS superconductor (dashed black line), as given by $S_\delta(T)=\textstyle\frac{S_\delta(0)\Delta(T)}{\Delta(0)}\tanh[\Delta(T)/2k_BT] $ \cite{Tinkham_2nd}. An independent way to extract $S_\delta$ from our TDTS measurements, without relying on any fits, is through the limit $S_\delta=\lim_{\omega\to0}\omega G_2$. We linearly extrapolate the measured $\omega G_2(\omega)$ down to $\omega=0$ (SM) and plot it on Fig.\ref{fig:zerofield}c to compare the two methods of determining $S_\delta$. As can be seen, there is good agreement between the two which validates our overall fitting procedure. 

\par 
The above analysis gives us important insights into the gap structure of the superconducting phase of Bi/Ni bilayer films. Some works (e.g.~\cite{Gong_chin_phylett_2015}) suggested that this system has a complex $p$-wave type gap structure which is naturally compatible with the observed TRS breaking, similar to what is believed to be realized in SrRu$_2 $O$_4$ \cite{Kapitulnik_PRL_2006,Luke_TRSB_1998}. Another possibility is complex $d$-wave pairing ($d_{xy}\pm id_{x^2-y^2}$), which is compatible with the surface crystal symmetry as argued by Gong et.~al.~\cite{Gong_Sci_ad_2017}. For these cases, the magnitude of the gap may be anisotropic and could lead to the observation of two energy gaps in the measurements of $G_1(\omega)$. However, our results on the Bi/Ni bilayer system closely correspond to those of a classic BCS weakly coupled  superconductor with a uniform gap. If two $p$ or $d$ wave components do exist then this implies that the system has an almost uniform gap structure with approximately equal magnitudes for each component \cite{Loder_srep_2015,Smidman_SOC_noncentrosymm_RPP_2017,RPSINGH_PRL_2014}. Such low anisotropy \cite{Gong_Sci_ad_2017} is consistent with the onset of superconductivity at a single transition temperature as observed.

\begin{figure*}
	\includegraphics[width=\textwidth,height=4.2cm]{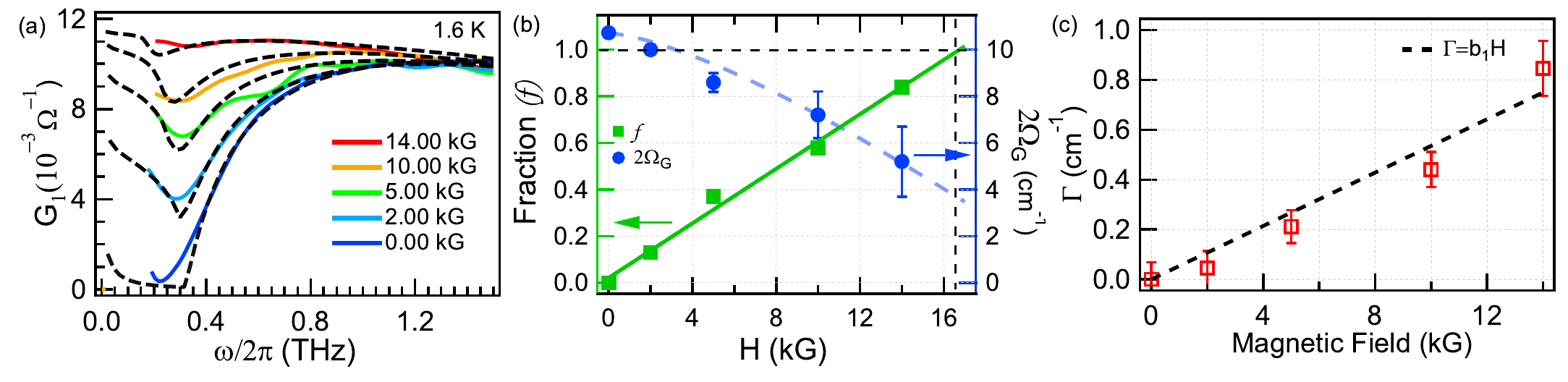}
	\caption{\label{fig:Magnet} (a) Out-of-plane field dependent real part, $G_1(\omega)$, (solid lines) of the complex conductance for Bi/Ni bilayer sample-C at $\SI{1.6}{\kelvin}$. The dashed lines are fits obtained by modeling the response within Maxwell Garnett theory, with the Drude model for the normal component and Mattis-Bardeen theory with effective spectroscopic gap  $\Omega_G$ for the superconducting component. (b) Field dependent $\Omega_G$ (blue dots) and the normal-volume fraction $f$ (green squares) with fit $f = H/H_{c2}$ (solid line). The dashed blue line is a guide to the eye. Horizontal and vertical dashed lines represent $f=1$ and $H_{c2}$ respectively. (c) Field dependent pair-breaking parameter $\Gamma$ fit to $\Gamma=bH$ (dashed line). The error bars represent the $\SI{95}{\percent}$ confidence interval.}
\end{figure*}

We now use TDTS measurements in both in-plane and out-of-plane magnetic fields to understand where the superconductivity develops in the Bi/Ni bilayer system. Fig.~\ref{fig:Parallel_MB}a,b shows $G_1(\omega)$ and $G_2(\omega)$ for a few in-plane magnetic fields at $T = \SI{1.6}{\kelvin}$ (see SM for data at other fields). The spectra show behavior similar to the zero-field temperature dependent spectra in Fig.~\ref{fig:zerofield}a, i.e, $ G_1(\omega)$ approaches its normal-state behavior with increasing magnetic field while the gap size reduces. Similar to the analysis above, we fit $ \tilde{G}(\omega)$ using Mattis-Bardeen theory with a single effective energy spectrum gap, $\Omega_G$ (dashed lines in Fig.~\ref{fig:Parallel_MB}a,b). We obtain reasonable fits for most of the frequency range but note that a small amount of spectral weight at low frequencies on this sample-B is not captured by the fits. This discrepancy is discussed below as possibly originating from disorder in the films. 

In general, the in-plane magnetic field results in pair breaking effects in the superconductor which leads to reduction in the pair-correlation gap, $ \Delta_p $. These effects can be quantified in terms of the Fermi velocity of the charge carriers through the behavior of the spectroscopic gap, $\Omega_G$, with field (Fig.~\ref{fig:Parallel_MB}b), as discussed below. This approach has also been applied for the electrodynamic response of NbN thin films~\cite{Tanner_PRL_2010}.  Here $\Omega_G$ can be related to the pair correlation gap, $\Delta_p$, via the relation $\Omega_G=\Delta_p[1-(\textstyle\frac{4}{\pi}\ln[\Delta_0/\Delta_p])^{2/3}]^{3/2}$ \cite{Maki_1964,Skalski_PR_1964}, where $\Delta_0 $ is the zero-field energy gap at $\SI{1.6}{K}$. The parameter $\Gamma$ which quantifies the strength of pair-breaking can then be found using the relation $\ln[\Delta_p/\Delta_0] = -\pi\Gamma/4\Delta_p$ for $\Gamma < \Delta_p$ \cite{Maki_1964,Parks_1969}.

The extracted values of $ \Omega_G $, $ \Delta_p $ and $ \Gamma $ as a function of in-plane field are shown in Fig.~\ref{fig:Parallel_MB}b. For a thin film superconductor in an in-plane magnetic field, $\Gamma$ is expected to be proportional to the square of the magnetic field $H$ \cite{Maki_1964,Parks_1969,Tinkham_2nd,Tanner_PRL_2010}, which is indeed the case here (Fig.~\ref{fig:Parallel_MB}b). An expression for $\Gamma$ in terms of the magnetic field is given by $\Gamma=bH^2 = D(eHd)^2/6$, where $d$ is the film thickness and $D =\uptau_{tr}\nu_f^2/3$ is the diffusion constant for a charge carrier at the Fermi level in terms of the transport collision time  $\uptau_{tr}$ and the Fermi velocity $\nu_f$ \cite{Tinkham_2nd,Parks_1969}. By fitting the pair-breaking parameter $\Gamma$ to $ bH^2 $ we obtain $b = \SI{0.0169\pm0.0009}{\centi\meter^{-1}\per T^2}$. Using $\uptau_{tr} = \SI{47.5e-14}{s}$ as determined from the Drude fit to the normal state and the film thickness $ d=\SI{11}{\nano\meter}$, we get $\nu_f = \SI{0.201\pm0.02e5}{\meter\per\second}$. This $ \nu_f $ is much smaller than Fermi velocities of all the orientations of Bi and Ni crystals (SM). This observation suggests that the superconducting quasi-particles do not belong to either of the individual components of the Bi/Ni bilayer separately. Note that in calculating $\nu_f$ we used the entire thickness of the Bi/Ni film ($d=\SI{11}{\nano\meter}$). Although in principle the effective thickness could be much less, this choice is further justified by the out-of-plane magnetic field dependence described below. 

In order to check the above determined value of $\nu_f$ without relying on the film thickness, we measure the optical response of the film to out-of-plane magnetic fields. In this case, the pair-breaking parameter is given by $ \Gamma = DeH $~\cite{Tinkham_2nd}. Fig.~\ref{fig:Magnet}a shows $G_1(\omega)$ for a number of out-of-plane magnetic fields at $T = \SI{1.6}{\kelvin}$. Note that this system is a type II superconductor and so an out-of-plane magnetic field above $H_{c1} \sim \SI{1.5}{\kilo\gauss}$ forms vortices with normal metal cores. As the wavelength of the probing THz beam is much greater than the size of the vortex cores ($\sim\SI{}{\nano\meter}$), and is at high frequencies, the resulting electrodynamic response can be modeled in terms of the Maxwell-Garnett theory (MGT) \cite{Garnett_1904} which is an effective medium theory. It has been applied to superconducting NbN thin films by Xi et.~al.~\cite{Tanner_PRB_2013}.
Within MGT, a superconducting thin film in an out-of-plane magnetic field is treated as a mixture of superconducting and normal metal components; where the superconducting component with volume fraction $ (1-f) $ is taken as the host medium and normal vortex cores  with volume fraction $ f $ as the embedded media \cite{Tanner_PRB_2013}. We again use Mattis-Bardeen theory, similar to the in-plane field data, to describe the superconducting component and the Drude model to describe the normal metal cores (see SM for full details on MGT). It is expected that due to the thin film geometry the magnetic field will almost uniformly penetrate the superconducting regions ($\Lambda_{\perp}=2\lambda^2/d = \SI{0.156}{mm})$.

The complex conductances for out-of-plane field dependent measurements are fit to the above described MGT using only $f$ and $\Omega_G $ as the free parameters. The resulting fits for $G_1(\omega)$ are shown as dashed lines in Fig.~\ref{fig:Magnet}a while the extracted values for $f$ and $\Omega_G $ as a function of magnetic field are plotted in Fig.~\ref{fig:Magnet}b. The volume fraction $f$ is related to the applied field as $f\sim H/H_{c2}$ \cite{Tanner_PRB_2013}, where $H_{c2}$ is the upper critical field. As can be seen in Fig.~\ref{fig:Magnet}b, $f \propto H$ and a simple linear extrapolation to $f=1$ yields $H_{c2} = 1.67 \pm\SI{0.19}{\tesla}$. This is in excellent agreement with the value of the upper critical field at $\SI{1.6}{\kelvin}$ determined from resistivity data ($\sim\SI{1.65}{\tesla}$) \cite{Gong_arxiv_2015} and so justifies our analysis of the electrodynamic response in terms of MGT.
Fig.~\ref{fig:Magnet}c shows the extracted values of $\Gamma$ as a function of field with a linear fit, $ \Gamma = b_1 H $, (dashed line) giving $ b_1 = \SI{0.536\pm0.005}{\centi\meter^{-1}\per T^2}$. Using $ b_1 = De $, we get $\nu_f = \SI{0.201\pm0.003e5}{\meter\per\second}$, which agrees with the value obtained from in-plane magnetic field data above. This confirms that the thickness of the superconducting film chosen in our earlier calculation is correct and it appears the entirety of the bilayer film becomes superconducting. 

Taken together with TRS breaking in the Bi/Ni bilayer, our observations of fully-gapped superconductivity occurring in the bulk of the system rather than just on the surface seem to suggest an effective $p_x \pm ip_y$ pairing symmetry as proposed in \cite{Chao_arxiv_2018}. Furthermore, given that above determined $\nu_f$ does not correspond to the Fermi-velocity of either Bi or Ni, and that the normal state conductance of the bilayer is significantly higher than either of pure Bi or Ni (SM), it is indeed likely that superconductivity originates in new states that occur due to formation of the bi-layer. Together with strong spin-orbit coupling from Bi and fluctuations from ferromagnetic Ni, this can lead to effective $ p $-wave like superconductivity \cite{Chao_arxiv_2018}.

\par Finally, we would like to discuss the discrepancy between the in-plane field data in Fig.\ref{fig:Parallel_MB}a,b and the Mattis-Bardeen type fits using a single gap. For a complex $ p $ or $ d $-wave order parameter, it is expected that an in-plane magnetic field may anisotropically suppresses one of the order parameter components preferentially giving a pure single component at some transition field below $H_{c2}$ (e.g., \cite{Agterberg_PRL_1998,Mao_PRL_2000}). This naturally results in low frequency absorption. It would be interesting to look for this transition field with other techniques such as heat capacity or NMR. Although the low frequency spectral weight we find may be reflective of this, another possibility is disorder in the films as they are highly susceptible to aging, air exposure and imperfections during growth. This disorder could lead to low frequency absorption and thus the fits underestimate $\tilde{G}(\omega)$ (e.g., \cite{Bing_PRB_2016,Swanson_PRX_2014}). We note that we can get better fits when we introduce a small Gaussian distribution in the gap as shown in SM (sec.VI) but these fits give roughly the same extracted parameters as above (SM). Thus, a small amount of disorder in this fashion does not affect our overall conclusions.

\begin{acknowledgments}
	Experiments at JHU were supported by the Army Research Office Grant W911NF-15-1-0560.  Film growth at Fudan was are supported by the National Basic Research Program of China (Grants No. 2015CB921402 and No. 2011CB921802), and the National Science Foundation of China (Grants No. 11374057, No. 11434003, and No. 11421404).
	
\end{acknowledgments}

\newpage
\setcounter{figure}{0}  
\renewcommand\thefigure{S\arabic{figure}} 
\renewcommand{\figurename}{Fig.} 

\renewcommand{\author}{}
\renewcommand{\title}{Supplementary Material:  Nodeless bulk superconductivity in the time-reversal symmetry breaking Bi/Ni bilayer system}
\begin{center}
	\textbf{\large Supplementary Material:  Nodeless bulk superconductivity in the time-reversal symmetry breaking Bi/Ni bilayer system}
\end{center}
\begin{center}
	\author{Prashant Chauhan$^1$,  Fahad Mahmood$^1$, Di Yue$^2$,  Peng-Chao Xu$ ^2 $, Xiaofeng Jin$ ^2 $  and N. P. Armitage$^1$\\
		\medskip
		$^1$ \textit{The Institute for Quantum Matter, Department of Physics and Astronomy, Johns Hopkins University, Baltimore, Maryland 21218, USA}\\
		$^2$ \textit{Department of Physics, Fudan University, Shanghai , China}}
\end{center}
\setlength\belowcaptionskip{-3ex}
\section{Zero-field DC conductance}
\begin{figure}[h]
	\includegraphics[width=7.5cm,height=4.5cm]{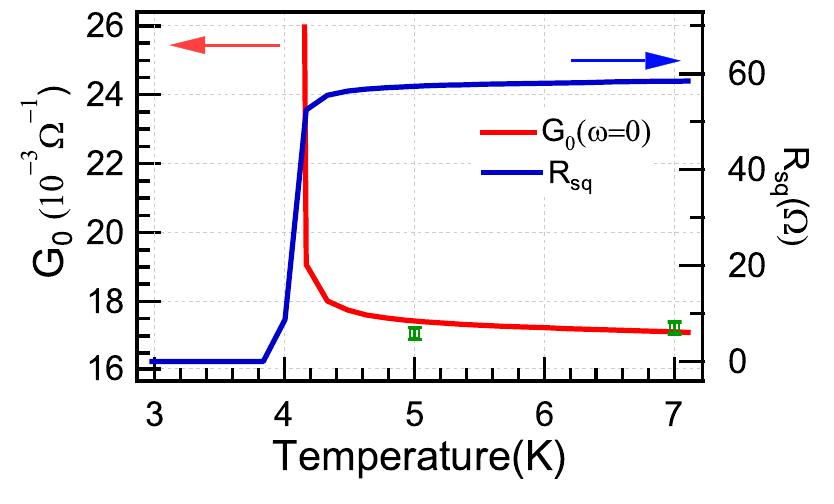}% Here is how to import EPS art
	\caption{\label{fig:resistance} Zero-field 2D sheet conductance ($ G_0 $) and sheet resistance (R$ _{sq} $) of the Bi/Ni bilayer. Green points are the $ \omega\rightarrow0 $  conductance extracted from fitting of the THz $ \tilde{G}(\omega) $ data with a Drude model. The error bars represent the $\SI{95}{\percent}$ confidence interval (2 s.d.).}
\end{figure}

The blue line in Fig.~\ref{fig:resistance} shows the temperature dependent zero-field four-probe 2D sheet resistance data taken on the Bi/Ni thin film sample. A single superconducting transition is seen at $ \sim\SI{4.15}{K} $. The 2D sheet conductance (red line) is obtained by inverting the resistance data. The green points in Fig.~\ref{fig:resistance} are the zero-field $ \omega\rightarrow0 $ conductance of the sample, obtained by fitting the normal state conductance spectrum at $\SI{5}{K}$ and $\SI{7}{K}$ using a Drude model (eq.~\ref{eq1}).

\section{Time-domain THz setup}
The complex conductance was obtained using time-domain THz spectroscopy. A femtosecond IR laser pulse is split along two paths to excite a pair of photoconductive `Auston'-switch antennae grown on LT-GaAs wafers. A broadband THz range pulse is emitted by one antenna and measured at the other antenna.  By varying the length-difference of the two paths, we map out electric field of the pulse as a function of time, both through the Bi/Ni sample on an MgO substrate and through a bare reference MgO substrate. The electric fields are converted to the frequency domain by taking a Fast Fourier Transform (FFT). By dividing the FFTs of the sample and reference scans, we	obtain the complex transmission of the sample. We then invert the transmission to obtain the complex conductance via the standard formula for thin films on a substrate:  $\tilde{\rm T}(\omega)=\frac{1+n}{1+n+Z_0\tilde{\sigma}(\omega)d} e^{i\Phi_s}$ where $\Phi_s$ is the phase accumulated from the small difference in thickness between the sample and reference substrates and $n$ is the substrate index of refraction. By measuring both the magnitude and phase of the transmission, both the real and imaginary conductance are obtained directly and so no Kramers-Kronig transformation is required. The complex conductance, ̃$\tilde{G}$, is then obtained from the complex transmission in the thin-film limit as $\tilde{G}\left(\omega\right)=\frac{(n+1)}{Z_0}(\frac{e^{iω(n-1)\frac{\delta L}{c}}}{T̃(\omega)} -1) $, where $n$ is the refractive index of the substrate (MgO) and $\delta L$ is the thickness difference between sample and reference substrates.

\section{{THz transmission in zero magnetic field}}
\begin{figure}
	\includegraphics[width=7.0cm,height=5.0cm]{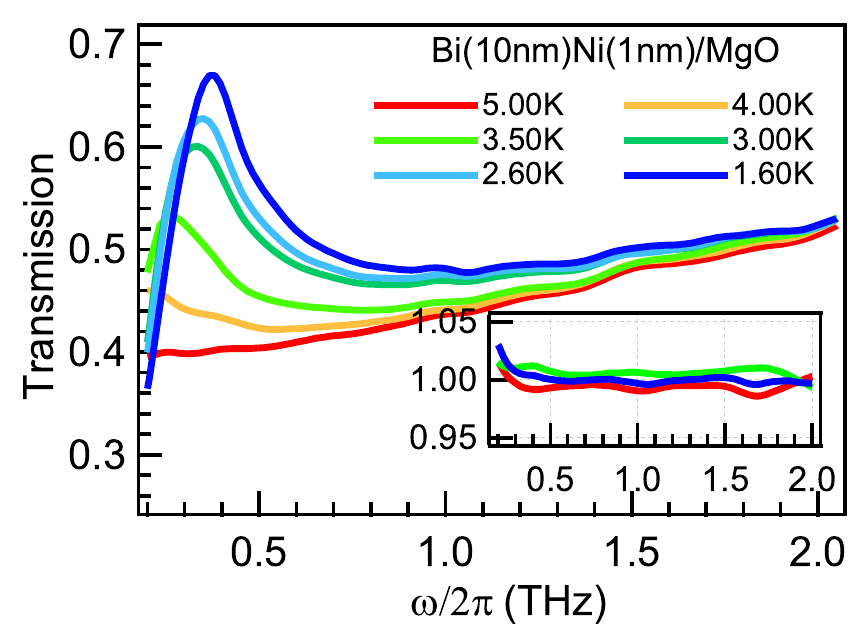}
	\caption{\label{fig:transmission} Transmission spectrum of the Bi/Ni bilayer in zero magnetic field. The inset shows $ 100\%$ lines for the MgO substrate, quantifying the uncertainty of data with respect to frequency.}
\end{figure}

Figure~\ref{fig:transmission} shows the magnitude of the frequency dependent complex transmission $ \tilde{\rm T}(\omega) $ of the Bi/Ni bilayer film at temperatures both above and below $\rm T_{c} $. Experimentally transmission is obtained as, $\tilde{\rm T}(\omega)= \rm \tilde{E}_{sam}(\omega)/\tilde{E}_{sub}(\omega) $, where $\rm \tilde{E}_{sam}(\omega) $ and $\rm \tilde{E}_{sub}(\omega) $ are the THz E-field measured at the receiver antennae for the sample and the substrate respectively. Below $\rm T_{c} $, we see a local maximum whose frequency and intensity increases with decreasing temperature. According to BCS theory, the frequency of this peak traces $ 2\Delta(\rm T) $, where $ \Delta(\rm T) $ is the temperature dependent superconducting gap.

\section{THz conductance in zero magnetic field}
\begin{figure}
	\includegraphics[width=7.0cm,height=9cm]{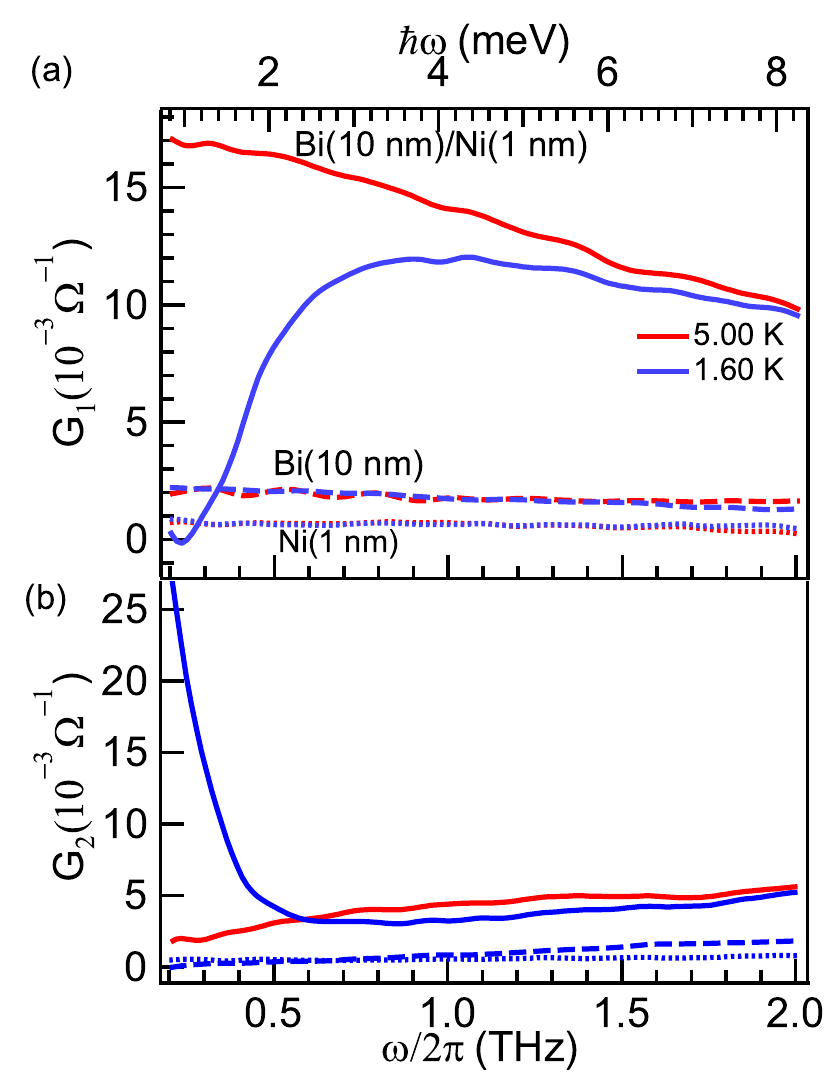}
	\caption{\label{fig:zf_G_compare} (a) Real and (b) imaginary parts of the zero-field conductance spectra from $ 0.2-\SI{2}{\tera\hertz} $  for the Bi/Ni bilayer thin film on an MgO substrate (solid lines) compared with the conductance of just Bi on MgO (dashed lines) and of just Ni on MgO (dotted lines) at $ \SI{1.6}{K} $.}
\end{figure}

In Fig.~\ref{fig:zf_G_compare} we compare the frequency dependent zero-field complex conductance of the Bi($\SI{10}{nm}$)/Ni($ \SI{1}{nm} $) bilayer thin film with that of just $ \SI{10}{nm} $ Bi(110)  and of just $ \SI{1}{nm} $ Ni(100) thin films. As can be seen, the Bi/Ni bilayer shows much higher conductance than that of the Bi film ($\sim 8 $ times greater) or of the Ni film ($\sim 16$ times greater). The conductance of the Bi/Ni bilayer is much higher than the sum of the conductances of its individual components. 

In the normal state ($\SI{5}{K}$) the real part of $\tilde{G}$, $G_1(\omega)$, for the Bi/Ni bilayer film shows a Drude-like Lorentzian peak feature while the imaginary part, $\tilde{G}_2(\omega)$ shows a positive dispersion corresponding to a finite scattering rate. We fit this normal state $\tilde{G}(\omega)$ with a Drude-Lorentz model description as follows: 
\begin{equation}\label{eq1}
\tilde{G}(\omega)=\epsilon_0 d\left(-\dfrac{\omega_p ^2}{i\omega-\Gamma_D}-i(\epsilon_\infty -1)\omega\right)
\end{equation}
where the first term in the bracket gives the Drude contribution and the second term gives the contribution to the dielectric constant from high frequency absorption. We also take into account the small contribution of $\epsilon_\infty$ as noted above in Eq.\ref{eq1}. The scattering rate determined for the sample used for data in Fig.\ref{fig:zf_G_compare} is $\Gamma_D =\SI{4.46}{THz}$ ($\uptau_{tr}= \SI{22.38e-14}{s} $). Note that the Bi/Ni bilayer has no residual conductance which might trivially be expected from its individual Bi or Ni layers. 

\section{In-plane magnetic field dependent THz conductance}
\begin{figure}
	\includegraphics[width=7cm, height=9cm]{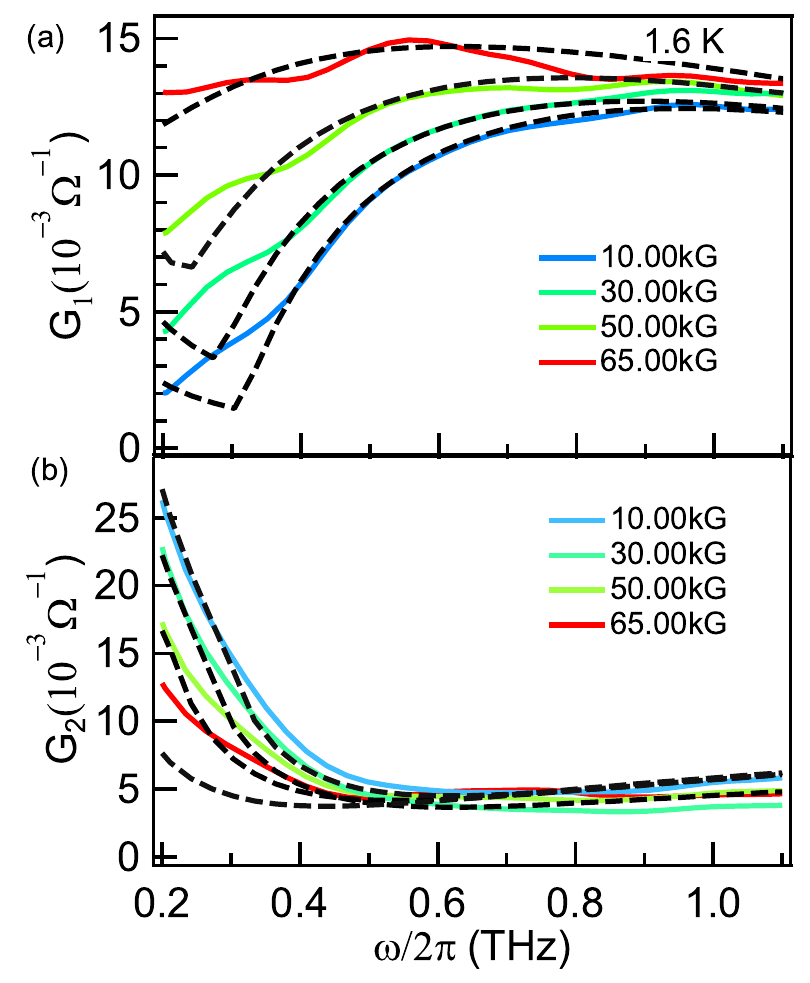}% Here is how to import EPS art
	\caption{\label{fig:img_cond_para} (a-b) In-plane magnetic field dependent real and imaginary parts of conductance, $\tilde{G}(\omega)$, (solid lines) for the Bi/Ni bilayer at $\SI{1.6}{K}$. The dashed lines are fits using Mattis-Bardeen theory (includes pair breaking effects) as described in the text. Data and fits for the other fields are shown in Fig.~2 of the main text.}
\end{figure}

Fig.\ref{fig:img_cond_para}a and b show the real $ G_1(\omega) $ and imaginary $ G_2(\omega) $ parts of the conductance respectively with in-plane magnetic fields. Fig.\ref{fig:img_cond_para}a shows data taken at four different fields. Data at four other fields is show in Fig.~2 of the main text. This data is fit to Mattis-Bardeen theory with a single effective energy spectrum gap $\Omega_G$ (dashed lines in Fig.\ref{fig:img_cond_para}a and b). 

\begin{figure}
	\includegraphics[width=7.cm,height=9.5cm]{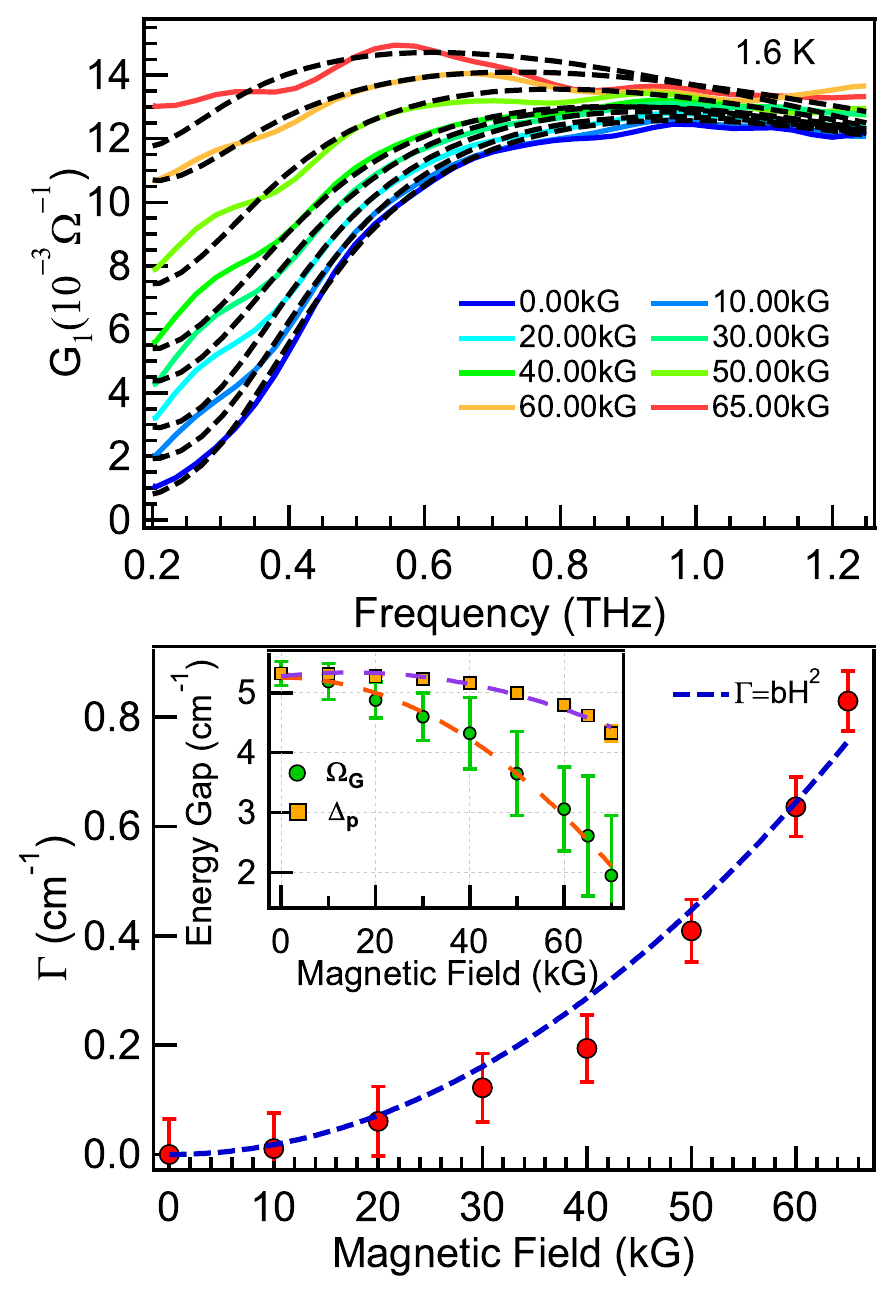}% Here is how to import EPS art
	\caption{\label{fig:real_cond_para} (a) In-plane magnetic field dependent real part of the optical conductance, $G_1(\omega)$, (solid lines) for the Bi/Ni bilayer at $\SI{1.6}{K}$ with fits (dashed lines) using Mattis-Bardeen theory with a Gaussian distribution for the effective spectroscopic gap. (b) Field dependence of the pair-breaking parameter $\Gamma$, determined from the optical conductance along with the fit $\Gamma = bH^2 $ (dashed line). Inset shows the field dependence of the effective spectroscopic gap $ \Omega_G $ (green dots) and the pair correlation gap $ \Delta_p$ (orange squares). The dashed red and purple lines are guide to the eye.}
\end{figure}

To better fit $\tilde{G}(\omega)$ at low frequencies, we model $G(\omega)$ similar to the way done in the main text but instead of using a single effective spectroscopic gap, $ \Omega_G $, we use a Gaussian distribution of the gap with a FWHM of $ \sim \SI{0.06 }{THz}$. Fig.\ref{fig:real_cond_para}a shows the resulting fits (dashed lines) to the data (solid lines). The values of $ \Omega_G $, $ \Delta_p $ and $ \Gamma $ extracted from the fits are similar to the ones obtained without using a Gaussian distribution. The fit $ \Gamma = bH^2 $ in Fig.\ref{fig:real_cond_para}b gives a value of $ b=\SI{0.0179}{cm^{-1}/T^2} $ from which we get Fermi velocity, $ \nu_f = \SI{0.210e5}{m/s} $. This value of $ \nu_f $ also agrees well with the Fermi velocity $ \nu_f = \SI{0.201e5}{m/s} $ obtained using the fits to the data with a single effective spectroscopic gap in the main text.

\section{Maxwell-Garnett Theory}
\begin{figure}
	\includegraphics[width=7.0cm,height=5.0cm]{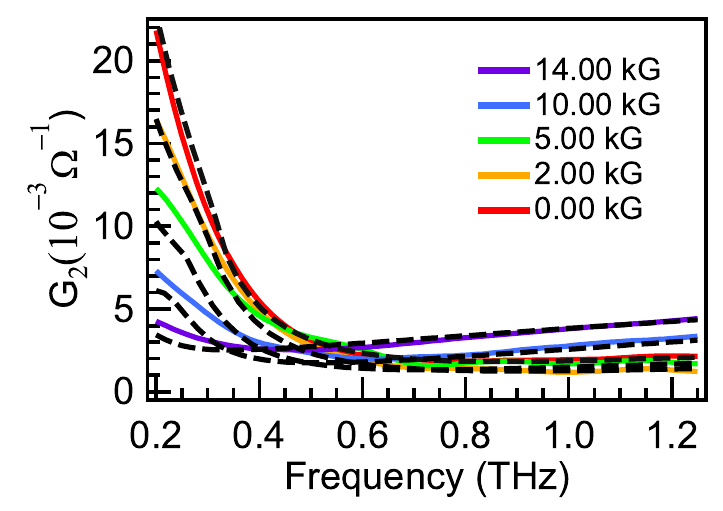}
	\caption{\label{fig:img_cond_perp} Out-of-plane magnetic field dependent imaginary part of the conductance, $G_2(\omega)$, (solid lines) of the Bi/Ni bilayer at $ \SI{1.6}{K} $. The dashed lines are simulated conductivity Maxwell-Garnett Theory fits using Mattis-Bardeen theory for the superconducting component (includes pair breaking effects) and Drude theory for the normal cores.}
\end{figure}

Maxwell-Garnett theory (MGT) can be used to describe the effective optical conductance, $ \tilde{G}(\omega) $, of a vortex-state superconductor. A superconducting thin-film in magnetic field, $ H >H_{c1} $, can be treated  as an inhomogeneous medium with two components, grain $ \textit{a} $ with volume fraction $ f $  embedded in a surrounding medium $ b $ with volume fraction $ 1-f $~\cite{Garnett_1904, Tanner_PRB_2013}. By assuming that the separation between the grains is large enough for an individual grain to scatter light and that the medium $b$ remains unaffected by grains, MGT gives an effective dielectric function for oriented ellipsoidal grains as~\cite{Tanner_PRB_2013,Carr_academic_1985,Granqvist_PRB_1977},
\begin{equation}
\tilde{\epsilon}_{MGT} = \tilde{\epsilon}_b + \tilde{\epsilon}_b\frac{f(\tilde{\epsilon}_a -\tilde{\epsilon}_b)}{g(1-f)(\tilde{\epsilon}_a -\tilde{\epsilon}_b) +\tilde{\epsilon}_b }
\end{equation} 
where $ g $ is the depolarization factor obtained from the shape of the ellipsoid. Taking vortices as cylindrical tubes with a normal core, we set $ g=1/2 $~\cite{Sihvola_Hindawi_JN_2007}. Using, $\tilde{\epsilon} = 1+4\pi i\tilde{G}/\omega  $, we obtain effective optical conductance as,
\begin{equation}
\tilde{G}(\omega) = \tilde{G_s} + \frac{2f(\tilde{G_N} -\tilde{G_s})(1+\dfrac{4\pi i\tilde{G_s}}{\omega})}{(1-f)(\dfrac{4\pi i}{\omega})(\tilde{G_N}-\tilde{G_s})+(1+\dfrac{4\pi i\tilde{G_s}}{\omega})}
\end{equation}
where, $f$ is the volume fraction of the normal metal cores, $ \tilde{G_N} $ and $ \tilde{G_s} $ are the conductances of the normal and superconducting fractions respectively.
In Fig.~\ref{fig:img_cond_perp} we show the imaginary part of the conductance, $ G_2(\omega) $, (solid lines) of the Bi/Ni bilayer. All the features of the data show good agreement with the MGT fits (dashed lines) for all fields.

\section{Fermi velocities of B\lowercase{i} and N\lowercase{i}}
\begin{table}[h]
	\caption{\label{tab:table1} Fermi velocities for different surfaces of rhombohedral Bi and Ni. We also quote the Fermi velocity obtained from our analysis in the last column.}\vspace{5mm}
	\begin{tabular}{|c|c|c|c|c|c|}
	\hline
	& \multicolumn{2}{c|}{{Bi(110)$^{\text{\tiny{\cite{Hofmann_NJP_2005}}}}$}} & \multicolumn{2}{c|}{Bi(100)$ ^{\text{\tiny{\cite{Hofmann_PRB_2005}}}} $}& {Bi$ ^{\text{\tiny{\cite{HOFMANN_PSS_2006}}}} $}\\
	\cline{2-5}
	$\nu_f(10^5$m/s)&{\fontsize{8}{5}\selectfont $\overline{M}-\overline{X}_2$} &{\fontsize{8}{5}\selectfont $\overline{\Gamma}-\overline{X}_2$} &{\fontsize{8}{5}\selectfont $\overline{\Gamma}_3-\overline{M}_3$}&{\fontsize{8}{5}\selectfont $\overline{\Gamma}_3-\overline{K}_3$}&(111)\\
	\cline{2-5}
	& 1.16 & 1.93 & 3.46 &5.64& 3.69 \\
	\hline
	\hline
	& Bi$ ^\text{{\tiny{\cite{Ashcroft}}}} $ & Ni $ ^\text{{\tiny{\cite{Ehrenreich_1963_PR}}}} $& Ni$ ^\text{{\tiny{\cite{Ehrenreich_1963_PR}}}} $ &Ni$ ^\text{{\tiny{\cite{Ehrenreich_1963_PR}}}} $&Expt\\
	
	$\nu_f(10^5$m/s)&Bulk&(110)&(100)&Bulk&Bi/Ni\\
	\cline{2-6}
	& 18.7 & 2.7 &3.9&10&0.20\\
	\hline
\end{tabular}
	\label{table:Bi_fermi_velocities}
\end{table}

\end{document}